\documentclass[graybox]{svmult}

\usepackage{mathptmx}       
\usepackage{helvet}         
\usepackage{courier}        
\usepackage{type1cm}        
\usepackage{makeidx}         
\usepackage{graphicx}        
\usepackage{multicol}        
\usepackage[bottom]{footmisc}

\makeindex             

\begin{document}

\title*{Emotional control - conditio sine qua non for
        advanced artificial intelligences?}
\titlerunning{Diffusive emotional control} 
\author{Claudius Gros}
\institute{Claudius Gros \at Institute for Theoretical Physics, 
           Goethe University Frankfurt, \email{gros07@itp.uni-frankfurt.de}}
%
%
\maketitle

\abstract*{Humans dispose of two intertwined information processing
           pathways, cognitive information processing via
neural firing patterns and diffusive volume control via neuromodulation.
The cognitive information processing in the brain is traditionally considered to
be the prime neural correlate of human intelligence, clinical studies
indicate that human emotions intrinsically correlate with the activation
of the neuromodulatory system.\newline\indent
We examine here the question: Why do humans dispose of the diffusive
emotional control system? Is this a coincidence, a caprice
of nature, perhaps a leftover of our genetic heritage, 
or a necessary aspect of any advanced intelligence, being it
biological or synthetic? \newline\indent
We argue here that emotional control is necessary to solve the
motivational problem, viz the selection of short-term utility functions,
 in the context of an environment where information, 
computing power and time constitute scarce resources.
          }

\abstract{Humans dispose of two intertwined information processing
           pathways, cognitive information processing via
neural firing patterns and diffusive volume control via neuromodulation.
The cognitive information processing in the brain is traditionally considered to
be the prime neural correlate of human intelligence, clinical studies
indicate that human emotions intrinsically correlate with the activation
of the neuromodulatory system.\newline\indent
We examine here the question: Why do humans dispose of the diffusive
emotional control system? Is this a coincidence, a caprice
of nature, perhaps a leftover of our genetic heritage, 
or a necessary aspect of any advanced intelligence, being it
biological or synthetic? \newline\indent
We argue here that emotional control is necessary to solve the
motivational problem, viz the selection of short-term utility functions,
 in the context of an environment where information, 
computing power and time constitute scarce resources.
          }

\section{Introduction}

The vast majority of research in artificial intelligences
is devoted to the study of algorithms, paradigms and 
philosophical implications of cognitive information
processing, like conscious reasoning and problem 
solving \cite{russell10}. Rarely considered is the 
motivational problem - a highly developed AI needs to 
set and select its own goals and tasks autonomously.

We believe that it is necessary to consider the 
motivational problem in the context of the observation
that humans are infused with emotions, possibly
to a greater extend than any other species \cite{dolan02}.
Emotions play a very central role in our lives, in literature 
and human culture in general. Is this predominance of
emotional states a coincidence, a caprice of nature, perhaps 
a leftover from times when we were still `primitives and
brutes', or perhaps a necessary aspect of any advanced intelligence? 

The motivational problem is about the fundamental conundrum 
that all living intelligences face. From the myriads of options
and behavioral strategies it needs to
select a single route of action at any given time.
These decisions are to be taken considering three 
limited resources, the information disposed of about the
present and the future state of the world, the time 
available to take the decision and the 
computational power of its supporting hard- or wetware. 
Here we argue that emotional control is deeply 
entwined with both short- and long-term decision 
making and allows to compute in real time 
approximate solutions to the motivational problem.

When considering the relation between
emotional control and the motivational problem
one needs to discuss the nature of non-biological 
intelligences for which this issue is of relevance.
We believe that, in the long term, there will be
two major developmental tracks in AI research -
focused artificial intelligences and organismic
universal synthetic intelligences. We believe that
the emotional control constitutes an inner core 
functionality for any universal intelligence 
and not a secondary addendum.

\section{Intelligent Intelligences}

We start with some terminology and a 
loose categorization of possible forms
of intelligence.

\runinhead{Focused Artificial Intelligences} 
We will use the term \textit{focused AI} for what constitutes
today's mainstream research focus in artificial 
intelligence and robotics.
These are highly successful and highly specialized
algorithmic problem solvers like the chess playing
program Deep Blue \cite{campbell99}, the DARPA-like 
autonomous car driving systems \cite{thrun06} and 
Jeopardy software champion Watson \cite{ferrucci10}.

Focused  artificial intelligences are presently the
only type of artificial intelligences suitable for
commercial and real-world applications. In the vast
majority of today's application scenarios a focused
intelligence is exactly what is needed, a reliable
and highly efficient solution solver or robotic controller.

Focused AIs may be able to adapt to changing demands and
have some forms of built-in, application specific learning
capabilities. They are however characterized by two
features.
\begin{itemize}
\item \underline{Domain specificity} A chess playing software
      is not able to steer a car. It is much more efficient to
      develop two domain specific softwares, one for chess and
      one for driving, than to develop a common platform.
\item \underline{Maximal a priori information} The performance
      real-world applications are generaly greatly boosted 
      when incorporating a maximal amount of a priori
      information into the architecture. Deep Blue contains the
      compressed knowledge of hundreds of years of human chess
      playing, the DARPA racing car software the Newton laws of
      motion and friction, the algorithms do not need to discover
      and acquire this knowledge from proper experiences.
\end{itemize}
Focused AI sees a very rapid development, increasingly driven
by commercial applications. They will become extremely powerful
within the next decades and it is questionable whether alternative
forms of intelligences, whenever the may be available in the
future, will ever be able to compete with focused AI on 
economical grounds. It may very well be, though difficult
to foretell, that focused AI will always yield a greater
return on investment than more general types of intelligences
with their motivational issues.

\runinhead{Synthetic Intelligences}
The term `artificial intelligence' has been used and abused
in myriads of ways over the past decades. It is standardly 
in use for mainstream AI research, or focused AI as described above.
We will use here the term \textit{synthetic intelligence} 
for alternative forms of intelligences, distinct from todays
mainstream route of AI and robotics research.

\runinhead{Universal Intelligences}
It is quite generally accepted that the human brain is an
exemplification of `universal' or `generic' intelligence.
The same wetware and neural circuitry can be used in many
settings - there are no new brain protuberances being formed
when a child learns walking, speaking, operating his fairy-tale
player or the alphabet at elementary school. There are parts of
the brain more devoted to visual, auditory or linguistic 
processing, but rewiring of the distinct incoming sensory data
streams will lead to reorganization processes of the respective 
cortical neural circuitry allowing it to adapt to new tasks and 
domains. 

The human brain is extremely adaptive, a skilled car driver
will experience, to a certain extend, its car as an extension
of his own body. A new brain-computer interference, when available
in the future, will be integrated and treated as a new sensory organ,
on equal footing with the biological pre-existing senses. 
Human intelligence is to a large extend not domain specific,
its defining trait is universality.

\runinhead{Organismic Intelligences}
An `organismic intelligence' is a real-world or simulated 
robotic system which has the task to survive. It is 
denoted \textit{organismic} since the survival task 
is generically formulated as the task to keep the 
support unit, the body, functional \cite{gros08,diPaolo03}.

Humans are examples of organismic intelligences. An
organismic synthetic intelligence may be universal, but 
not necessarily. The term `organismic' is not to 
be confused with `embodiment'. Embodied AI 
deals with the question whether considering the
physical functionalities of robots and bodies is helpful,
of even essential, for the understanding of cognitive 
information processing and intelligence in general
\cite{anderson03,pfeifer07,froesea09}.

\runinhead{Cognitive System}
The term `cognitive system' is used in various ways in
the literature, mostly as a synonym for a cognitive 
architecture, viz for an information processing 
domain-specific software. I like to reserve the term 
\textit{cognitive system} for an intelligence which 
is both universal and organismic, may it be biological 
or synthetic. 

Humans are biological cognitive systems in this sense
and most people would expect, one can however not foretell
with certainty, that `true' or `human level AI' would 
eventually be realized as synthetic cognitive systems.
It is an open and unresolved questions, as a matter of 
principle, whether forms of human level AI which are not
cognitive systems in above sense, are possible at all. 

\runinhead{Human Level Artificial Intelligences}
An ultimate goal of research in artificial and
synthetic intelligences is to come up with 
organizational principles for intelligences of
human or higher level. How and when this goal will
be achieved is presently in the air, a few aspects
will be discussed in the next section. This has not
precluded an abundance of proposals on how to test
for human-level intelligences, like the Turing test
\cite{turing50}
or the capability to perform scientific research. 
Some people believe that human intelligence will have
been achieved when we do not notice it.

\runinhead{The Complexity Conundrum}
\begin{figure}[t]
\sidecaption[t]
\includegraphics[width=0.55\textwidth]{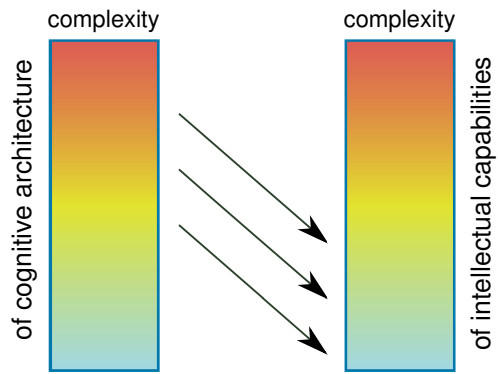}
%
%
\caption{Illustration of the (hypothetical) complexity conundrum,
which regards the speculation that the mental capabilities of 
biological or synthetic intelligences (right) might be systematically
too low to fully understand the complexity of their own 
supporting cognitive architectures (left). In this case the
singularity scenario would be void.
}
\label{fig:complexity-conundrum}       
\end{figure}

Regarding the issue when and how humanity
will develop human level intelligences we
discuss here shortly the possible occurance
of a `complexity paradox', for which we will
use the term \textit{complexity conundrum}.

Every intelligence arises form a highly organized
soft- or wetware. One may assume, though this
is presently nothing more than a working
hypothesis, that more and more complex brains
and software architectures are needed for higher
and higher intelligences. The question is than,
whether a brain with a certain degree of complexity
will give raise to a level on intelligence capable
to understand its own wetware, 
compare Fig.~\ref{fig:complexity-conundrum}.  
It may be, as a matter
of principle, that the level of complexity a certain
level of intelligence is a able to handle is always
below the level of complexity of its own supporting
architecture.

This is really a handwaving and rather philosophical
question with many open ends. Nevertheless one may
speculate whether the apparent difficulties of present-day
neuroscience research to carve out the overall working 
principles of the brain may be in part due to a complexity
conundrum. Equivalently, considering the successes and
the failures of over half a century of AI research,
our present near-to complete ignorance of the
overall architectural principles necessary for the development
of eventual human level AI may be routed similarly in either 
a soft or a strong version of the complexity conundrum.

The complexity conundrum would however not, even if true, 
preclude humanity to develop human level artificial or synthetic 
intelligences in the end. As a last resort one may proceed 
by trial and error, viz using evolutionary algorithms, or via
brute force reverse engineering, if feasible. The notion of
a complexity conundrum is relevant also to the popular concept of a
singularity, a postulated runaway self improving circle 
of advanced intelligences \cite{vinge93,chalmers10}. 
The complexity conundrum, if existing in any form,
would render the notion of a singularity void, 
as it would presumably apply to intelligences at all levels.

\section{Routes to Intelligence}

There are presently no roadmaps, either individually
proposed or generally accepted, for research and
development plans leading to the ultimate goal
of highly advanced intelligences. Nevertheless there
are two main, conceptually distinct, approaches.

\begin{figure}[t]
\sidecaption[t]
\centerline{
\includegraphics[width=0.90\textwidth]{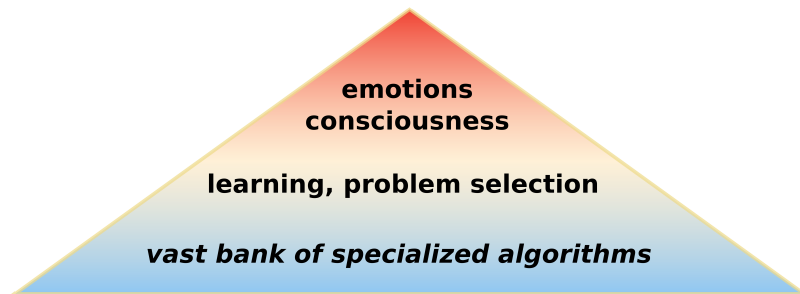}
           }
%
%
\caption{Mainstream architecture for a hypothetical human-level
artificial intelligence. The motivational problem would be delegated
to a secondary level responsible of selecting appropriate
modules for problems and tasks which are not autonomously generated
but presumably presented to the AI by human supervisors. Higher
cognitive states like consciousness are sometimes postulated to emerge 
spontaneously with raising complexity from self-organizational principles,
emotional control is generically regarded as a later-stage add-on, if at all.}
\label{fig:pyramid_AI}       
\end{figure}

\subsection{From focused to general intelligence?}
The vast majority of present-day research efforts
is devoted to the development of high-performing 
focused intelligences. It is to be expected that we 
will see advances, within the next decades, along 
this roadmap for hundreds and many more application 
domains.

There is no generally accepted blueprint on how to go beyond
focused intelligences, a possible scenario is presented in
Fig.~\ref{fig:pyramid_AI}. A logical next step would be to
hook up a vast bank of specialized algorithms, the focused
intelligences, adding a second layer responsible for switching
between them. This second layer would then select the algorithm 
most appropriate for the problem at hand and could contain
suitable learning capabilities. 

This kind of selection layer constitutes a placebo for the
motivational problem, the architecture presented in 
Fig.~\ref{fig:pyramid_AI} would not be able to 
autonomously generate its own goals. This is however 
not a drawback for industrial and for the vast
majority of real-world applications, for which the artificial
intelligence is expected just to efficiently solve problems and
tasks presented to it by human users and supervisors.

In a third step it is sometimes expected that cognitive 
architectures may develop spontaneously consciousness 
with raising levels of complexity. This speculation,
particularly popular with science-fiction media, is
presently void of any supporting or 
contrarian scientific basis \cite{tononi98,koch99}.
Interesting is the tendency of mainstream AI to discuss
emotions as secondary features, mostly useful to
facilitate human-robot interactions \cite{handbookEmotions09}.
Emotions are generically not attributed a central role in
cognitive architectures withing mainstream AI.

One could imagine that the kind of cognitive architecture 
presented in Fig.~\ref{fig:pyramid_AI} approaches,
with the expansion of its basis of focused intelligences,
step by step the goal of a universal intelligence able 
to handle nearly any conceivable situation. It is unclear 
however which will be the pace of progress towards this goal. 
It may be that progress will be initially very fast, 
slowing then however down substantially when artificial 
intelligence with elevated levels of intellectual 
capabilities have been successfully developed. This kind of
incremental slowing-down is not uncommon for the pace of
scientific progress in general. Life expectancy
has been growing linearly, to give an example, over the last
two centuries. The growth in life expectancy is extremely steady
and  still linear nowadays, despite very rapidly 
growing medical research efforts. Not only in economics, but
also in science there are generic decreasing returns on growing investments.
Similarly, vast increases in the number and in the power of
the underlying array of focused intelligences may, in the end,
lead to only marginal advances towards universality.

\begin{figure}[t]
\sidecaption[t]
\centerline{
\includegraphics[width=0.90\textwidth]{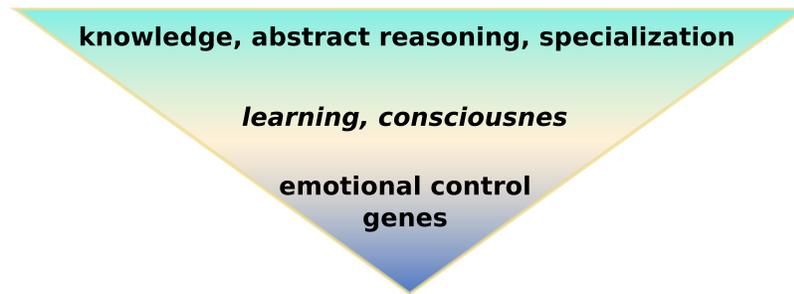}
           }
%
%
\caption{Architecture for biologically inspired universal
synthetic intelligences, viz of cognitive systems. The basis
would be given by a relatively small number of
genetically encode universal operating principles,
with emotional control being central for the further 
development through self-organized learning processes.
How consciousness would arise in this setting is not
known presently, it is however regarded as a prerequisite
for higher intellectual capabilities such as abstract
reasoning and knowledge specialization.
}
\label{fig:pyramid_CS}       
\end{figure}

\subsection{Universal learning systems}

The only real-world existing example of an
advanced cognitive system is the mammalian brain.
It is hence reasonable to consider biologically 
inspired cognitive architectures. Instead of
reverse engineering the human brain, one tries then
to deeply understand the general working principles
of the human brain. 

There are good arguments that self-organization
and general working principles are indeed dominant
driving forces both for the development of the brain
and for its ongoing functionality 
\cite{kohonen82,haken08}. Due to the small
number of genes in the human genome, with every 
gene encoding only a single protein, direct genetic 
encoding of specific neural algorithms has either to
be absent all together in the brain or be limited 
to only a very small number of vitally important 
features.

It is hence plausible that a finite number of working
principles, possibly as small as a few hundred, 
may be enough for a basic understanding of the 
human brain, with higher levels of complexity 
arising through self-organization.  Two examples 
for general principles are `slowness' \cite{foldiak91}
for view-invariant object recognition and 
`universal prediction tasks' \cite{gros08}
for the autonomous generation of abstract concepts.

Universality, in the form of operating principles,
lies therefore at the basis of highly developed cognitive
systems, compare Fig.~\ref{fig:pyramid_CS}. This is
in stark contrast to mainstream AI, where universality
is regarded as the long-term goal, to be reached when
starting from advanced focused intelligences.

One of the genetically encoded control mechanisms at
the basis of a cognitive system is emotional control,
which we will discuss in more detail in the next
section. Emotional control is vitally important for
the functioning of a universal learning system, and not
a secondary feature which may be added at a later stage.
\begin{itemize}
\item \underline{Learning} In the brain two dominant 
      learning mechanisms are known. Hebbian-type
      synaptic plasticity which is both sub-conscious
      and automatic, and reward-induced learning,
      with the rewards being generated endogenously
      through the neuromodulatory control system, the
      later being closely associated with the experience
      of motions.
\item \underline{Goal selection} 
      Advanced cognitive systems are organismic and hence
      need to constantly select their short- and long term
      goals autonomously, with emotional weighing of
      action alternatives playing a central role.
\end{itemize}
It is not a coincidence, that the emotional control
system is relevant for above two functionalities,
which are deeply inter-dependent. There can be no
efficient goal selection without learning from
successes and failure, viz without reward induced
learning processes.

\begin{figure}[t]
\sidecaption[t]
\centerline{
\includegraphics[width=0.90\textwidth]{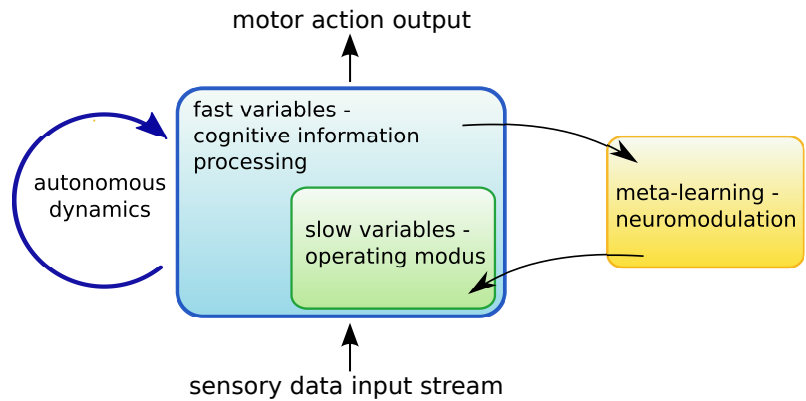}
           }
%
%
\caption{Fast and slow variables have distinct
functionalities in the brain, with the operating
modus (mood) being set by the slow variables and the
actual cognitive processes, which are either input 
induced or autonomous \cite{gros09,gros10}, 
being performed by the fast variables. 
The adaption of the slow variables (metalearning)
is the task of the diffusive neuromodulatory system
(emotional control). 
}
\label{fig:metalearning}       
\end{figure}

\section{Emotional Control}

Emotions are neurobiologically not yet
precisely defined. There are however substantial
indications from clinical studies that emotions 
are intrinsically related to either the tonic or
the phasic activation of the neuromodulatory 
system \cite{fellous99}. For this reason we 
will denote the internal control circuit involving 
neuromodulation, compare Fig.~\ref{fig:metalearning},
\textit{emotional control}. We will also use the 
expression \textit{diffusive emotional control} 
since neuromodulation acts as a diffusive volume effect.

\begin{svgraybox}
One needs to differentiate between the functionality
of emotions in the context of cognitive system theory, 
discussed here, and the experience (the qualia) of emotions.
It is presently an open debate whether the body is 
necessary for the experience of emotions and moods, 
which may be induced by the proprioceptual sensing
of secondary bodily reactions \cite{barrett07}. The 
origin of emotional experience is not subject 
of our deliberations.
\end{svgraybox}

\subsection{Neuromodulation and metalearning}

Animals dispose of a range of operating modi,
which one may identify with moods or emotional
states. A typical example of a set of
two complementary states is exploitation vs.\
exploration: When exploitive the animal is
focused, concentrated on a given task and
decisive. In the explorative state the animal
is curious, easily distracted and prone to
learn about new aspects of his environment.
These moods are induced by the tonic,
respectively the phasic activation of the
neuromodulatory system \cite{krichmar08},
the main agents being Dopamine, Serotonin,
Norepinephrine and Acetylcholine.

When using the language of dynamical system theory
we can identify the task of the neuromodulatory system
with metalearning \cite{doya02}. Any complex system
disposes of processes progressing on distinct
time scales. There may be in principle a wide range
of time scales, the simplest classification
is to consider slow and fast processes driven 
respectively by slow and fast variables.

Cognitive information processing is performed in 
the brain through neural firing and synaptic 
plasticity, corresponding to the fast variables in 
terms of dynamical system theory \cite{gros08}. The 
general operating modus of the neural circuitry, 
like the susceptibility to stimuli,
the value of neural thresholds or the pace of
synaptic plasticities are slow degrees of freedom.
The adaption of slow degrees of freedom to changing
tasks is the realm of metalearning, which in the brain
is preformed through the neuromodulatory system,
compare Fig.~\ref{fig:metalearning}.

Metalearning is a necessary component of any complex
dynamical system and hence also for any evolved synthetic 
or biological intelligence. It is therefore not surprising that the
human brain disposes of a suitable mechanism. Metalearning
is also intrinsically diffusive, as it involves the modulation
not of individual slow variables, metalearning is about
the modulation of the operating modus of entire dynamical
subsystems. It is hence logical that the metalearning
circuitry of the brain involves neuromodulatory neurons
which disperse their respective neuromodulators, when activated,
over large cortical or subcortical areas, modulating the
behavior of downstream neural populations in large 
volumes.

An interesting and important question regards the guiding
principles for meta\-learning. An animal has at its disposal
a range of distinct behaviors and moods, foraging, social
interaction, repose, exploration, and so on. Any cognitive
system is hence faced with a fundamental time allocation 
problem, what to do over the course of the day. The 
strategy will in general not be to maximize time allocation
of one type of behavior, say foraging, at the expense of
all others, but to seek an equilibrated distribution of
behaviors. This guiding principle of metalearning has been
denoted `polyhomeostatic optimization' \cite{markovic10}.

\subsection{Emotions and the motivational problem}

It is presently unclear what distinguishes metalearning
processes which are experienced as emotional from
those which are unconscious and may hence be termed `neutral'.
It has been proposed that the difference may be that
emotional control has a preferred level of activation,
neutral control not \cite{gros09Emotions,gros10Emotions}. 
When angry one generally tries behavioral strategies aimed 
at reducing the level of angriness and internal rewards 
are generated when successful. In this view emotional 
control is intrinsically related to behavior and learning, 
in agreement with neuro-psychological observations 
\cite{krichmar08,dolan02,baumeister07}.

Emotional states induce, quite generically, problem
solving strategies. The cognitive system either
tries to stay in its present mood, in case 
it is associated with positive internal rewards,
or looks for ways to remove the causes for its 
current emotional state, in case it is associated 
with negative internal rewards. Emotional control
hence represents a way, realized in real-world 
intelligences, to solve the motivational problem,
determining the utility function the intelligence
tries to optimize at any given point of time.

A much discussed alternative to emotional control
is straightforward maximization of an overall
utility function \cite{marcus05}. This paradigm is highly 
successful when applied to limited and specialized
tasks, like playing chess, and is as such important 
for any advanced intelligence. Indeed we argue that
emotional control determines the steady-state utility
function. As an example consider playing chess. Your
utility function may either consist in trying to beat 
the opponent chess player or to be defeated by your
opponent (in a non-so-evident way) when playing together 
with your son or daughter. These kinds of utility functions
are shaped in real life by our emotional control 
mechanisms.

It remains however doubtful whether it would be 
possible to formulate an overall, viz a long-term utility 
function for a universal intelligence and to compute in 
real time its gradients. Even advanced hyper-intelligences 
will dispose of only an exponentially small knowledge about 
the present and the future state of the world, prediction
tasks and information acquisition is
generically NP-hard (non-polynomial)
\cite{chickering94,nikoloski08,sieling08}.
Time and computing power (however large it may be)
will forever remain, relatively seen, scarce resources.
It is hence likely that advanced artificial
intelligences will be endowed with `true' synthetic
emotions, the perspective of a hyper-intelligent robot
waiting emotionless in its corner, until its human
boss calls him to duty, seems implausible
\cite{arbib04,ziemke09,parisi10}.

Any advanced intelligence needs to be a twofold 
universal learning system. The intelligent system
needs to be on one side able to acquire any kind
of information in a wide range of possible
environments and on the other side to determine
autonomously what to learn, viz solve the
time allocation problem. The fact that both facets
of learning are regulated through diffusive 
emotional control in existing advanced intelligences
suggests that emotional control may be a conditio
sine qua non for any, real-world or synthetic,
universal intelligence.

\section{Hyper-emotional trans-human intelligences?}

Looking around at the species on our planet one
may surmise that increasing cognitive capabilities
go hand in hand with rising complexity and
predominance of emotional states \cite{dolan02}.
The rational is very straightforward. An animal with
say only two behavioral patterns at its disposition,
e.g.\ sleeping and foraging, does not need
dozens of moods and emotions, in contrast to
animals with a vast repertoire of potentially
complex behaviors.

This observation is consistent with the theory
developed here, that metalearning as a diffusive
emotional control system is a necessary component
for any synthetic and biological intelligence.
It is also plausible that the complexity 
the metalearning control needs to increase
adequately with increasing cognitive capacities.

It is hence amusing to speculate, whether synthetic
intelligences with higher and higher cognitive
capabilities may also become progressively emotional.
Super-human intelligences would then also be
hyper-emotional. An outlook in stark contrast to
the mainstream view of hyper-rational robots, 
which presumes that emotional states will be 
later-stage addendums to high performing artificial 
intelligences.

\begin{acknowledgement}
I acknowledge lively discussions and feedback at 
the conference on the Philosophy and Theory of 
Artificial Intelligence, PT-AI, 
Thessaloniki, October 3-4 (2011).
\end{acknowledgement}



\end{document}